\documentstyle[11pt]{article}
\input epsf
\begin{document}

\thispagestyle{empty}
\begin{center}

\medskip
\vspace{2.5 cm} {\large \textbf{GELFAND-DICKEY ALGEBRA AND HIGHER SPIN \\ SYMMETRIES ON $T^2=S^1\times S^1$}}\\
\vspace{2 cm} \textbf{M.B. SEDRA}\footnote{Associate of ICTP: sedra@ictp.it},\\
{\small \ \ \ } {\small \ Universit\'{e} Ibn Tofail, Facult\'{e}
des
Sciences, D\'{e}partement de Physique,}\\

{\small \ \ \ }{\small \ Laboratoire de Physique de La Mati\`ere
et Rayonnement (LPMR), K\'{e}nitra, Morocco}\\

{\small \ Groupement National de Physique de Hautes Energies,
GNPHE, Morocco,}\\

{\small \ Abdus Salam International Centre for Theoretical
Physics,
Trieste, Italy.}\\

\end{center}
\vspace{0.5cm} \centerline{\bf Abstract} \baselineskip=18pt
\bigskip
We focus in this work to renew the interest in higher conformal
spins symmetries and their relations to quantum field theories and
integrable models. We consider the extension of the conformal
Frappat et al. symmetries containing the Virasoro and the
Antoniadis et al. algebras as particular cases describing
geometrically special diffeomorphisms of the two dimensional torus
$T^2$. We show in a consistent way, and explicitly, how one can
extract these generalized symmetries from the Gelfand-Dickey
algebra. The link with Liouville and Toda conformal field theories
is established and various important properties are discussed.
\\

\hoffset=-1cm \textwidth=11,5cm \vspace*{1cm}
\hoffset=-1cm\textwidth=11,5cm \vspace*{0.5cm}
\newpage

\newpage
\section{Introduction} For several years conformal field
theories and their underlying Virasoro algebra \cite{CFT} have
played a pioneering role in the study of string theory
\cite{string}, critical phenomena in certain
statistical-mechanical models \cite{stat} and integrable systems
\cite{int}. The conformal symmetry is generated by the stress
energy momentum tensor of conformal spin $2$ whose short-distance
operator product expansion (OPE) is given by
\begin{equation}
T(z)T({\omega})=\frac{c/2}{(z-\omega)^{4}}+\frac{2T({\omega})}
{(z-\omega)^{2}}+\frac{\partial T({\omega})}{(z-\omega)}+...
\end{equation}
Some years ago, Zamolodchikov opened a new issue of possible
extensions of the Virasoro algebra to to higher conformal spins.
Among his results, was the discovery of the $W_3$ algebra
involving, besides the usual spin-$2$ energy momentum tensor
$T(z)$, a conformal spin-$3$ conserved current \cite{zamo1,
zamo2}. $W_3$-symmetry is an infinite dimensional algebras
extending the conformal invariance (Virasoro algebra) by adding to
the energy momentum operator, $T(z)\equiv W_{2}$, a set of
conserved currents $W_{s}(z)$ of conformal spin $s>2$ with some
composite operators
necessary for the closure of the algebra. \\\\
In the language of $2d$ conformal field theory, the above
mentioned spin-$3$ currents of the $W_3$-symmetry are taken in
general as primary satisfying the short-distance operator product
expansion (OPE) \cite{zamo1, zamo2}
\begin{equation}
\begin{array}{lcl}
T(z)W(\omega)&=&\frac{3 W(\omega)}{(z-\omega)^{2}}+\frac{\partial
W(\omega)}{(z-\omega)},\\\\
W(z)W(\omega)&=&\frac{c/3}{(z-\omega)^{6}}+\frac{2
T(\omega)}{(z-\omega)^{4}}+\frac{\partial
T(\omega)}{(z-\omega)^{3}}+\frac{1}{(z-\omega)^2}[2\beta\Lambda(\omega)+\frac{3}{10}\partial^{2}T(\omega)]\\\\
&+&\frac{1}{(z-\omega)}[\beta\partial\Lambda(\omega)+\frac{1}{15}\partial^{3}T(\omega)]\\\\
\end{array}
\end{equation}
where
$\Lambda(\omega)=(TT)(\omega)-\frac{3}{10}\partial^{2}T(\omega)$
and $\beta=\frac{16}{22+5c}$. This symmetry which initially was
identified as the symmetry of the critical three Potts model, has
also been realized as the gauge symmetry of the so-called $W_3$
gravity \cite{Wgrav}. Since then higher spin extensions of the
conformal algebra have been studied intensively by using different
methods: field theoretical \cite{f1, f2, f3}, Lie algebraic
\cite{lie} or geometrical approach \cite{ga}. $W_3$-algebras are
also known to arise from the Gelfand-Dickey bracket (second
Hamiltonian structure) of the generalized Korteweg-de Vries (KdV)
hierarchy \cite{GD, GD1, GD2}. In this context, the correspondence
is achieved naturally in terms of pseudo-differential Lax
operators, ${\mathcal L}_{n}=\sum u_{n-j}\partial^{j}$, allowing
both positive as well as nonlocal powers of the differential
$\partial ^{j}$ where the spin-$j$ currents $u_j$ are satisfying
integrable non linear differential equations. The integrability of
these equations can be traced to the fact that KdV hierarchy's
equations are associated to higher conformal spin symmetries
through the
Gelfand-Dickey Poisson bracket.\\\\
Another kind of higher conformal spin symmetries deals with the
Frappat et al. algebra \cite{frap} on the bidimensional torus
$T^2$. This is generated by the infinite dimensional basis set of
mode operators $L_{k,l}$, $k, l\in Z$ verifying the classical
commutation relations
\begin{equation}
[L_{k, l}, L_{r, s}]=[m_{0}(k-r)-n_{0}(l-s)]L_{k+r, l+s}
\end{equation}
where $n_0$ and $m_0$ are two arbitrary parameters. The above
algebra has the property of offering a unified definition of
several extensions of the two dimensional conformal algebra. In
fact it's shown that the standard Virasoro algebra as well as the
Antoniadis et al. algebras \cite{anton} are special cases of
eq(3). Later on, a $W_3$-extended conformal symmetry going beyond
the Frappat et al. algebra, on the torus $T^2$, is proposed in
\cite{sss} by adding the spin-3 currents. These several extended
conformal symmetries are important as they are associated to
higher diffeomorphisms namely diff($T^2$) extending the standard
diffeomorphisms, diff($S^{1}$), involved in the $2d$ conformal
invariance.\\\\
Note by the way that, contrary to the case of fractional
superconformal symmetries, the Frappat et al. symmetries as well
as their $W_3$ extensions consist in replacing the manifold $S^1$
by a higher dimensional compact manifold ${\cal M}$. This can
offer a large framework for studying more general extensions of
$2d$-conformal symmetries including fractional superconformal
symmetries and their generalizations
\cite{frac}.\\\\
The principal goal of this work concerns the study of some non
standards properties of higher conformal spin symmetries in the
bidimensional torus $T^2$. This is motivated, in one hand, by the
increasingly important role that play the infinite-dimensional Lie
algebras in the development of theoretical physics. Best known
examples are given by the Virasoro algebra, which underlies the
physics of $2d$ conformal field theories (CFT) and its
$W_k$-extensions. On the other hand, it's today well recognized
that $2d$ conformal symmetry and it's $W_k$ higher spin extensions
are intimately related to the algebra of diff$(S^1)$ and
diff$(T^2)$ respectively. Therefore, we know that $2d$-conformal
symmetry and the KdV integrable model are intimately related in
the standard diff$(S^1)$ framework. Such relation is shown to
arise from the second hamiltonian structure of the KdV integrable
system reproducing then the classical form of the Virasoro
algebra.\\\\
One wonder whether the area preserving diffeomorphisms on $T^2$
can be related to integrable hierarchy systems in the same spirit.
In this context, and after a setup of our conventional notations
and basic definitions, we develop a systematic analysis leading to
an explicit derivation of the Gelfand-Dickey Poisson bracket. This
is based on the theory of pseudo-differentials operators
characterized by the following diff$(T^2)$-non standard Lax
differential operators $\{log H,
\}^{n}+\sum\limits_{i=0}^{n-2}u_{n-i}(z, \omega)\{log H, \}^{i}$,
where the derivation is given by the hamiltonian vector field
$\xi_{H}\equiv \{log H, \}$ playing in diff$(T^2)$ the same role
of the derivation $\partial_{z}$ in diff$(S^1)$. The particular
KdV-like Lax differential operator $\xi_{H}^{2}+u_2$ associated to
the KdV equation is compatible with the conformal-invariant
Liouville-like differential equation $\{log {\bar{K}}, \{log H,
\phi\}\}=e^{\phi}$. The compatibility is assured once the current
$u_{2}(z, \omega)$, shown to satisfy the following KdV-like non
linear differential equation $\frac{\partial u_{2}}{\partial
t_{3}}=\frac{3}{2}u_2\{log H, u_2\}-\frac{3}{4}\{log H,
u_2\}^{(3)}$, behaves under a bicomplex change of coordinates like
the stress energy momentum tensor on the bidimensional torus $T^2$
namely ${\mathcal{T}}(z, \omega)=-\{log H, \phi \}^{(2)}-\{log H,
\phi \}^2$. Several important properties are discussed with some
concluding remarks at the end.
\section{Bianalytic fields on the torus $T^2$}
In this section we give the general setting of the basic
properties of the algebra of bianalytic fields defined on the
bidimensional torus $T^2$. \\\\
{\bf 1)} We start first by noting that the ring $\mathcal R$ of
conformal fields defined on the bidimensional torus $T^2$, and
transforming as primary fields under a special subalgebra of
diff$(T^2)$ , can be thought of as a generalization of the ring of
the Virasoro primary fields of the $2d$ real space
${\mathbf{R}}^2\approx \mathbf{C}$. The two dimensional torus
$T^2$ is viewed as a submanifold of the $4d$ real space
${\mathbf{R}}^4\approx {\mathbf{C}}^2$. It is parametrized by two
independent complex variables $z$ and $\omega$ and their
conjugates $\bar z$ and $\bar \omega$ satisfying the constraint
equation ${z\bar z}=\omega \bar \omega =1$. Solutions of these
equations are given by $z=e^{i n\theta}, \omega=e^{im\psi}$ where
$n$ and $m$ are two integers and where $\theta$ and $\psi$ are two
real parameters.\\\\
{\bf 2)} We identify the ring ${\cal R}$ of bianalytic fields on
$T^2$ to be ${\cal R}\equiv \widehat{\Sigma }^{(0,0)}$ the tensor
algebra of bianalytic fields of arbitrary conformal spin. This is
a an infinite dimensional $SO(4)$ Lorentz representation that can
be written as
\begin{equation}
\widehat{\Sigma }^{(0,0)}=\oplus_{k \in Z} \widehat{\Sigma }_{(k,
k)}^{(0,0)}
\end{equation}
where the $\widehat{\Sigma }_{(k, k)}^{(0,0)}$'s are one
dimensional $SO(4)$ irreducible modules corresponding to functions
of bianalytic conformal spin $(k, k)$. The upper indices $(0, 0)$
carried by the spaces figuring in eq(4) are special values of
general indices $(p, q)$ to be introduced later on. The generators
of $\widehat{\Sigma }_{(k, l)}^{(0,0)}$ are biperiodic arbitrary
functions that we generally indicate by $f(z, \omega)$. These are
bianalytic functions expanded in Fourier series as
\begin{equation}
f(z,\omega )=\sum\limits_{n,m\in Z}f_{nm} z^{n} \omega ^{m}
,\hspace{1cm} \partial_{\bar z} f=\partial_{\bar \omega}f=0
\end{equation}
where the constants $f_{nm}$ are the Fourier modes of $f$. This is
nothing but a generalization of the usual Laurent expansion of
conformal fields on the complex plane $\mathbf{C}$. Note by the
way that the integers $n$ and $m$ carried by the Fourier modes
$f_{nm}$ are nothing but the $U(1)\times U(1)$ Cartan charges of
the $SO(4)\approx SU(2)\times SU(2)$ Lorentz group of the Euclidan
space $\mathbf{R}^4$. Bianalytic functions on $\mathbf{C}^2$
carrying $U(1)\times U(1)$ charges $r$ and $s$ as follows
\begin{equation} f_{{(r,s)}} (z,\omega )=\sum\limits_{n,m\in
Z}f_{nm} z^{n-r} \omega ^{m-s};, \hspace {1cm} r,s\in Z,
\end{equation}
such that the eq.(5) appears then as a particular example of
eq(6). Note also that we can define the constants on $\mathcal{R}$
as been ${f}_{(0,0)}$ such that
\begin{equation}
\partial _{z, \bar z}{ f}_{0,0}=0=\partial _{\omega, \bar\omega}{{f}_{(0,0)}}.
\end{equation}
The coefficients $f_{nm}$ are given by
\begin{equation}
f_{mn} =\oint\limits_{c_{1} } \frac{dz}{2i\pi }z^{-n-l+r}
\oint\limits_{c_{2} } \frac{d\omega }{2i\pi } \omega ^{-m+-l+s}
f_{(r,s)}(z,\omega ),
\end{equation}
where $c_{1}\times c_{2}$ is the contour surrounding the
singularity $(z,\omega)=(0,0)$ in the complex space.
\\\\
{\bf 3)} The special subset $\widehat{\Sigma}_{(k,
k)}^{(0,0)}\subset {\cal R}$ is generated by bianalytic functions
$f_{(k, k)}$, $k \geq 2$. They can be thought of as the higher
spin currents involved in the construction of the $W$-algebra on
$T^2$ \cite{sss}. As an example, the following fields
\begin{equation}
\begin{array}{lcl}
W_{(2, 2)}&=&{u}_{(2, 2)}(z, \omega)\\
W_{(3, 3)}&=&{u}_{(3, 3)}(z, \omega)-\frac{1}{2}\{log H, {u}_{(2,
2)}\}
\end{array}
\end{equation}
are shown to play the same role of the spin-$2$ and spin-$3$
conserved currents of the Zamolodchikov $W_3$ algebra \cite{zamo1,
zamo2}. Next we will denote, for simplicity, the fields $u_{(k,
k)}(z, \omega)$ of conformal spin $(k, k)$, $k\in Z$ simply as
$u_{k}(z, \omega)$. The function log($H$) will be also described
later.\\\\
{\bf 4)} The Poisson bracket on $T^2$ is defined as follows
\begin{equation}
\{f,g\}=\partial_{z}f\partial_{\omega}g-\partial_{z}g\partial_{\omega}f
\end{equation}
with $\{z,\omega \}=1$. We denote $\{f, .\}=\xi_{f}$ and
$\xi_{f}.g=\{f, .\}.g=\{f, g\}+g\{f, .\}$ equivalently this shows
how the Poisson bracket on the torus can play the role of a
derivation. For convenience we will adopt the following notation
$\xi_H\equiv\xi_{log H}$ as been the hamiltonian vector field
operator associated to the arbitrary function $H$. This
hamiltonian operator is shown, in the present context of $T^2$, to
replace the standard derivation $\partial_z$ describing the basis
of differential operators on the circle $S^1$. We have also
\begin{equation}
\{log H, .\}\circ f\equiv\xi_{H}.f=\{log H, f\}+f\xi_H
\end{equation}
showing a striking resemblance with action of $S^1$-differential
operators ,
\begin{equation}
\partial f= f'+f\partial; \hspace {1cm} f'\equiv (\partial f)
\end{equation}
Later on the differential operators that will be used are simply
integer powers of $\xi_H$ subject to the following rule
\begin{equation}
[\xi_{f}, \xi_{g}]=\xi_{[f, g]}
\end{equation}\\
{\bf 5)} We introduce new form of the differential operators
associated to the symmetry of $T^2$. Before that we have to
consider the following conventions notations \cite{GD2}:%
\newline
\textbf{a)} $\widehat{\Sigma }_{n}^{(r,s)}$: This is the space of
differential operators of conformal weight $(n, n)$ and degrees
$(r,s)$ with $ r\leq s$. Typical operators of this space are given
by
\begin{equation}
\sum_{k=r}^{s}{u}_{n-k}(z, \omega)\xi_{H}^{k}
\end{equation}
\newline
This is the analogue of diff$(S^1)$ operators
\begin{equation}
\sum_{k=r}^{s}{u}_{n-k}\partial^{k}
\end{equation}
\textbf{b)} $\widehat{\Sigma }_{n}^{(0,0)}\subset {\cal R}$: This
space describes the conserved currents $u_{n}(z, \omega)$.\\
\textbf{c)} ${\widehat{\Sigma }}_{n}^{(k,k)}$: Is the space of
differential operators type,
\begin{equation}
u_{n-k}(z, \omega)\xi_{H}^{k}.
\end{equation}
\newline
\textbf{d)} The residue operation: ${Res}$
\begin{equation}
{Res}(f\xi_{H}^{-1})=f.
\end{equation}
\\{\bf 6}. To summarize, we give here bellow a table
containing the essential properties of the mathematical objects
involved in diff$(T^2)$-higher conformal spin symmetries analysis
\begin{equation}
\begin{tabular}{ccc}\\\\
\textbf{Objects $\mathcal{O}$} &\hspace{1cm} \textbf{The conformal
weight $|\mathcal{O}|$}
\\\\
$z$, $\omega$, $\partial_{z}$, $\partial_{\omega}$  & $|z|=(-1, 0)$, $|\omega|=(0, -1)$, $|\partial_{z}|=(1, 0)$, $|\partial_{\omega}|=(0, 1)$ \\ \\
$L_{k,l}$ & $|L_{k,l}|=(-k, -l)$ \\ \\
${\mathcal T}(z,\omega)\equiv {W}_{2}(z,\omega)$ & $|{\mathcal T}(z,\omega)|=(2, 2)$ \\ \\
$W_{s}(z,\omega), s=2,3,...$ & $|W_{s}(z,\omega)|=(s, s)$ \\\\
$\{f,g\}^{(k)}=\underbrace{ \{f, \{f,{...}\{f } \limits_{k}, g\}$
& \hspace{0,5cm} $|\{f,g\}^{(k)}|= (k, k) +k|f|+|g|$ \\\\
$\{f,g\}^{k}=\underbrace {\{f, g\}{...}\{f, g\}}\limits_{k}$
& \hspace{0,5cm} $|\{f,g\}^{k}|= (k, k) +k|f|+k|g|$ \\\\
$\xi_{H}=\{log H, .\}$ & \hspace{0,5cm} $|\xi_H|= (1, 1)$\\\\
$Res$ & $|Res|=(1, 1)$
\end{tabular}
\end{equation}

\section{Higher conformal spin symmetries on $T^2$}

\subsection{The Frappat et al. conformal symmetry}

This is the algebra generated by the infinite dimensional basis
set of mode operators $L_{k,l}, k, l\in Z$ satisfying the
classical commutation relations (without central extension)
\begin{equation}
[L_{k, l}, L_{r, s}]=[m_{0}(k-r)-n_{0}(l-s)]L_{k+r,l+s}
\end{equation}
where $n_{0}$ and $m_{0}$ are two arbitrary parameters. The above
algebra has the property of offering a unified definition of
several extensions of the $2d$ conformal algebra. Here with some
particular examples:\\\\
\textbf{1. The Virasoro algebra}\\
Setting for instance $m_{0}=1$ and $n_{0}=l=s=0$, we get the
classical Virasoro algebra
\begin{equation}
[L_k, L_r]=L_{k+r}+\frac{c}{12}k(k-1)\delta_{k+r, 0}
\end{equation}
for vanishing central charge $c=0$.\\\\
\textbf{2. The Antoniadis et al algebra}\\
Setting now $m_{0}=s$ and $n_{0}=r$ one recover the Antoniadis et
al algebra given by
\begin{equation}
[L_{k, l}, L_{r, s}]=[ks-lr]L_{k+r,l+s}
\end{equation}

\subsection{The $W_{3}$ generalization}

Since the Frappat et al. algebra is relevant in describing
geometrically special diffeomorphisms of the two dimensional torus
$T^{2}$ and extend successfully the known symmetries namely the
Virasoro and the Antoniadis ones, we focussed in previous works
\cite{sss} to go beyond this extension and improve it much more.
In fact the central charge as well as the $W_{3}$-extension of the
Frappat et al. algebra on $T^{2}$ are built. The first important
contribution of \cite{sss} consists in computing the missing
central charge of the Frappat et al. algebra, we find
\begin{equation}
c(k,l,m_0,n_0)=\frac{c}{12}\prod_{j=0,\pm}[m_{0}(k-j)-n_{0}(l-j)]\delta_{k+r}\delta_{l+s}
\end{equation}
The second important contribution deals with the derivation of the
$W_{3}$-extension of the Frappat et al. algebra. This is generated
by the mode operators $L_{k,l}$ and $W_{k,l}$ and reads as
\begin{equation}
\begin{array}{lcl}
[L_{k,l}, W_{r,s}]&=&[m_{0}(2r-k)-n_{0}(2s-l)]W_{k+r,l+s}\\

[W_{k,l}, W_{r,s}]&=&[m_{0}(k-r)-n_{0}(l-s)](\alpha
L_{k+r,l+s}+\beta \Lambda_{k+r,l+s})\\
&+&\frac{c}{360}\prod_{j=0,\pm
1}[m_{0}(k-j)-n_{0}(l-j)]\delta_{k+r}\delta_{l+s}\\
\end{array}
\end{equation}
where $\beta =\frac{16}{22+5c}$ and $\alpha=\alpha_1+\alpha_2$
with
\begin{equation}
\begin{array}{lcl}
\alpha_1&=&\frac{1}{12}[m_{0}(k+r+3)-n_{0}(l+s+3)][m_{0}(k+r+2)-n_{0}(l+s+2)],\\\\
\alpha_2&=&\frac{-1}{6}[m_{0}(k+2)-n_{0}(l+2)][m_{0}(r+2)-n_{0}(s+2)],\\\\
\end{array}
\end{equation}
and where $\Lambda _{k+r, l+s}$ is a composite operator given by
\begin{equation}
\Lambda _{k+r,l+s}= \sum_{i,j\in
Z}(:L_{i,j}L_{k+r-i,l+s-j}:-\frac{3}{10}\alpha_{1}L_{k+r,l+s})
\end{equation}
The important remark at this level is that this derived
diff$(T^2)$ extended $W_3$-algebra contains the well known
diff$(S^1)$ Zamolodchikov $W_3$ algebra \cite{zamo1} as a special
subalgebra obtained by setting $n_0=0$ and $m_0=1$. This new
symmetry eqs.(23-25) which extend also the Antoniadis et al.
algebra eq.(21) offers moreover a unified definition of several
generalizations of the two dimensional $W_3$ algebra exactly as do
the Frappat et al. algebra with respect to the Virasoro
symmetry.\\\\
In the OPE language, we introduce a bianalytic conserved current
${\mathcal T}(z,\omega)$ defined on the bicomplex space
$\mathbf{C}^2$ parametrized by the complex variables $z$ and
$\omega$ and their conjugates $\bar z$ and $\bar \omega$. The
particular field ${\mathcal T}(z,\omega)$, generalizing the $2d$
energy momentum tensor expands in Laurent series as
\begin{equation}
{\mathcal T}(z,\omega)=\sum_{k,l\in
Z}z^{-k-2}\omega^{-l-2}L_{k,l};\hspace{1cm}
\partial_{\bar z}{\mathcal T}=\partial_{\bar \omega}{\mathcal T}=0
\end{equation}
or equivalently
\begin{equation}
L_{m,n} =\oint\limits_{c_{1} } \frac{dz}{2i\pi }z^{m+1}
\oint\limits_{c_{2} } \frac{d\omega }{2i\pi } \omega ^{n+1}
{\mathcal T}(z,\omega ),
\end{equation}%
where $c_{1}\times c_{2}$ is a contour surrounding the singularity
$(z,\omega)=(0,0)$ in the complex space. The OPE analogue of the
extended Virasoro algebra eqs.(19) and (22) is given by \cite{sss}
\begin{equation}
{\mathcal T}(1){\mathcal T}(2) = \frac{c}{12} \left\{ Log
H,\frac{1}{\omega _{12} z_{12} } \right\} ^{(3)} +2\left\{ Log
H,\frac{1}{\omega _{12} z_{12} } \right\}
{\mathcal T}(2)+\left\{ LogH,{\mathcal T}(2)\right\} \frac{1}{\omega _{12} z_{12} } \\
\end{equation}
where we have set ${\mathcal T}(k)={\mathcal T}(z_{k},
\omega_{k})$ for short and where $H=H(z, \omega)$ is an arbitrary
bianalytic function in $\mathbf {C}^ 2$. A particular choice of
this arbitrary function leading to eqs.(19) and (22) is given by
$H(z, \omega)=z^{-n_0}\omega^{-m_0}$.\\\\
Note that the symbol \{,\} appearing in eq.(28) is the usual
Poisson bracket defined, for any pair of bianalytic functions
$f(z, \omega)$ and $g(z, \omega)$, as
\begin{equation}
\{f,
g\}=\partial_{z}f\partial_{\omega}g-\partial_{\omega}f\partial_{z}g
\end{equation}
with the following property
\begin{equation}
\{f, g\}^{(3)}=\{f,\{f,\{f, g\}\}\},
\end{equation}
Note moreover that in establishing eqs.(19) and (22) from the OPE
eq.(28) one finds the following expression
\begin{equation}
\begin{array}{lcl}
[L_{k, l}, L_{r, s}]&=&((s-l)\frac{\partial log H}{\partial log
z}-\frac{\partial log H}{\partial log \omega}(r-k))L_{k+r,l+s}\\\\
&+&\frac{c}{12}\prod_{j=0,\pm}((k-j)\frac{\partial log H}{\partial
log z}-(l-j)\frac{\partial log H}{\partial log
\omega})\\\\
\end{array}
\end{equation}
The $W_3$ extended symmetry is generated by a conserved current
$W(z,\omega )$ of conformal weight $h_{z,\omega}=(3, 3)$. This is
simply seen at the level of the energy-momentum tensor ${\mathcal
T}(z, \omega)$ who transforms under the bianalytic coordinate
change in $\mathbf{C}^2$
\begin{equation}
z\rightarrow \tau (z, \omega), \hspace{1cm} \omega \rightarrow
\sigma (z, \omega)
\end{equation}
as follows
\begin{equation}
{\mathcal T}(z, \omega)=(\partial_{z}
\tau)^{2}(\partial_{\omega}\sigma)^{2}{\mathcal \widetilde{T}}(z,
\omega)+\frac{c}{12}{\mathcal S}(z, \omega, \tau, \sigma)
\end{equation}
where ${\mathcal S}(z, \omega, \tau, \sigma)$ is the Schwartzian
derivative given by
\begin{equation}
{\mathcal S}(z, \omega, \tau, \sigma)=\{log H, log
(\partial_{z}\tau\partial_{\omega}\sigma)\}^{(2)}-\frac{1}{2}[\{log
H, log (\partial_{z}\tau\partial_{\omega}\sigma)\}]^2
\end{equation}
One easily observe that the conformal current $\mathcal T$
exhibits a conformal weight $h_{z,\omega}=(2,
2)$.\\
In the OPE language , the $W_3$ extension of the FRSTH algebra
read in addition to the eq.(28) as
\begin{equation}
\begin{array}{lcl}
{\mathcal T}(1)W(2) & = & 3\left\{ log H,\frac{1}{\omega _{12}
z_{12} } \right\} W(2)+\left\{ log H,W(2)\right\} \frac{1}{\omega
_{12} z_{12} } , \\
W(1)W(2) & = & \frac{c}{360} \left\{ log H,\frac{1}{\omega _{12}
z_{12} } \right\} ^{(5)} +\frac{1}{3} \left\{ log
H,\frac{1}{\omega _{12} z_{12} }
\right\} ^{(3)} {\mathcal T}(2) \\
& & +\frac{1}{15\omega _{12} z_{12} } \left\{ log H,{\mathcal T}(2)\right\} ^{(3)} +%
\frac{1}{3} \left\{ log H,\frac{1}{\omega _{12} z_{12} } \right\}
^{(2)}
\left\{ log H,{\mathcal T}(2)\right\} \\
& & +\frac{3}{10} \left\{ log H,\frac{1}{\omega _{12} z_{12} }
\right\}
\left\{ log H,{\mathcal T}(2)\right\} ^{(2)} \\
& & +2\beta \left\{ log H,\frac{1}{\omega _{12} z_{12} } \right\}
\Lambda (2)+\beta \left\{ log H,\Lambda (2)\right\}
\frac{1}{\omega _{12} z_{12} } ,
\end{array}
\end{equation}
where
\begin{equation}
\begin{array}{lcl}
W(z, \omega)& = & \sum_{k,l\in Z}z^{-k-3}\omega^{-l-3}W_{k,l};\\
\Lambda(z, \omega)&=&\sum_{k,l\in
Z}z^{-k-4}\omega^{-l-4}\Lambda_{k,l};
\end{array}
\end{equation}

\subsection{Diff$(T^2)$ Conformal transformations} Consider an
arbitrary finite and bianalytic coordinate change:
\begin{equation}
\begin{array}{lcl}
z\rightarrow\tilde z&=&\sigma(z, \omega)\\
\omega\rightarrow\tilde \omega&=&\tau(z, \omega)
\end{array}
\end{equation}
Under this symmetry, the functions $f_{(r,s)}(z, \omega)$
transform as:
\begin{equation}
f_{(r,s)}(z,
\omega)=(\partial_{z}\sigma)^{r}(\partial_{\omega}\tau)^{s}\tilde{f}_{(r,s)}(\sigma,
\tau)
\end{equation}
For an infinitesimal variation
\begin{equation}
\begin{array}{lcl}
\tilde{z}&=& z-B\\
\tilde{\omega}&=& \omega+A\\
\end{array}
\end{equation}
where $A(z, \omega)$ and $B(z, \omega)$ are two arbitrary
bianalytic functions of the $U(1)\times U(1)$ charges $(0, -1)$
and $(-1, 0)$ the same as $\omega$ and $z$ respectively, we have
\begin{equation}
\delta f_{(r,s)}(z, \omega)=V(A, B)
f_{(r,s)}+(s\partial_{\omega}A-r\partial_{z}B)f_{(r,s)}
\end{equation}
where the vector field $V(A, B)=A\partial_{\omega}-B\partial_{z}$,
obeys the Lie algebra of diffeomorphisms diff($T^2$) namely
\begin{equation}
[V(A_1, B_1), V(A_2, B_2)]=V(A, B)
\end{equation}
with
\begin{equation}
\begin{array}{lcl}
A&=&(A_{1}\partial_{\omega} A_{2}-B_{1}\partial_{z} B_{2})-(1\leftrightarrow2)\\
B&=&(A_{1}\partial_{\omega} B_{2}-B_{1}\partial_{z}
A_{2})-(1\leftrightarrow2)
\end{array}
\end{equation}
In the case of diff$(S^1)$, corresponding to set $A=0$ and $B\neq
0$ (resp. $B=0$ and $A\neq 0$) the functions $f_{(r,s)}$ behaves
as a two dimensional conformal objects of weight $h=r$ ( resp.
$h=s$). However the situation, in which both $A$ and $B$ are non
vanishing independent bianalytic functions, leads to see
$f_{(r,s)}$ as a conformal field of diff$(S^1)\times$ diff$(S^1)$.
Note however that diff$(T^2)$ is a special case of
diff$(S^1)\times$ diff$(S^1)$ corresponding for example to set
$r=s$ at the level of eq(40) and choose the functions $A$ and $B$
to transform as
\begin{equation}
\begin{array}{lcl}
A&=&{\tilde A}(\frac{\partial_{z}H}{H})\\\\
B&=&{\tilde A}(\frac{\partial_{\omega}H}{H})
\end{array}
\end{equation}
where $\tilde A$ and $H(z, \omega)$ are respectively arbitrary
bianalytic functions of $U(1)\times U(1)$ charges $(-1, -1)$ and
$(n_{0}, m_{0})$ with $n_{0}, m_{0}\in \mathbf{Z}$. Note by the
way that setting $\tilde{A}=H$, one discovers the area preserving
diffeomorphisms algebra of the torus studied in \cite{anton,
arak}. Further, by imposing the constraint
\begin{equation}
\partial_{\omega}H=0 \hspace{1cm} but\hspace{1cm} \partial_{z} H\neq 0,
\end{equation}
and vice versa $\partial_{\omega }H\neq 0$ but $\partial_{z}H =
0$, the equations reduce to a conformal variation of one of the
two Diff(S$^{1}$) subgroups of diff(T$^{2}$). We shall consider
hereafter the case ($\partial_{z}H)(\partial_{\omega }H)\neq 0$.
With the choice eq(40), eq.(43) takes the following remarkable
form :
\begin{equation}
\delta f_{r} =\tilde{A}\left\{ \log H, f_{r} \right\} +r\left\{
\log H,\tilde{A}\right\} f_{r} ,
\end{equation}
where we have set $f_{r}=f_{(r,r)}$ for short and where we have
used the Poisson bracket defined as:
\begin{equation}
\left\{\log H,G\right\} =(\partial_{z} \log H)\partial_{\omega }
G-(\partial_{\omega } \log H)\partial_{z} G.
\end{equation}
It is interesting to note here that eq.(45) exhibits a striking
resemblance with the conformal transformation of a two-dimensional
primary field of conformal weight h $=$ r. This is why we refer to
the set ${\cal R}$ of bianalytic functions transforming like
eq.(45) as been the ring of conformal fields on the torus T$^{2}$.
Elements of this ring are then primary fields on C$^{2}$ defining
highest weight representations of the conformal algebra on the
torus.

\section{Higher spin symmetries from the Gelfand-Dickey analysis}
We describe here the basic features of the algebra of arbitrary
differential operators, refereed hereafter to as ${\widehat
\Sigma}$, acting on the ring $\mathcal {R}$ of analytic functions
on $T^2$. We show in particular that any such differential
operator is completely specified by a conformal
weight\footnote{For a matter of simplicity we denote objects
$X_{(n, n)}$ of conformal weight $|X_{(n, n)}|=(n, n)$ simply as
$X_{n}$} $(n, n)$, $n\in Z$, two integers $r$ and $s$ with
$s=r+i$, $i\geq 0$ defining the lowest and the highest degrees,
respectively, and finally $(1+r-s)=i+1$ analytic fields $u_{j}(z,
\omega)$. Its obtained by summing over all the allowed values of
spin (conformal weight) and degrees in the following way:
\begin{equation}
{\widehat{\Sigma }}=\oplus _{r\leq s}\oplus _{n\in Z}{\widehat{%
\Sigma }}_{n}^{(r,s)}.
\end{equation}
with ${\widehat{\Sigma }}_{(n, n)}\equiv {\widehat{\Sigma }}_{n}$.
Note that the space ${\widehat{\Sigma }}$ is an infinite
dimensional algebra which is closed under the Lie bracket without
any condition. A remarkable property of this space is the
possibility to introduce six infinite dimensional classes of
sub-algebras related to each other by special duality relations.
These classes of algebras are given by ${\widehat{%
\Sigma }}_{s}^{\pm }$, with $s=0,+,-$ describing respectively the
different values of the conformal spin which can be zero, positive
or negative. The $\pm$ upper indices stand for the sign (positive
or negative) of the degrees quantum numbers, for more details see
\cite{GD2}.

\subsection{Setup of the GD integrable analysis on $T^2$}
\subsubsection{ The space ${\widehat \Sigma}_{n}^{(r,s)}$}

To start let's precise that the space ${\widehat \Sigma}
_{n}^{(r,s)}$ contains differential operators of fixed conformal
spin $(n, n)$ and degrees (r,s), type
\begin{equation}
{\cal L}_{n}^{(r,s)}(u)=\sum_{i=r}^{s}u_{n-i}(z,
\omega)\circ\xi_{H}^{i},
\end{equation}
These are $\xi_H$'s polynomial differential operators extending
the hamiltonian field $\xi_{H}=\{log H, \}$. Elements ${\cal
L}_{n}^{(r,s)}(u)$ of ${\widehat \Sigma} _{n}^{(r,s)}$ are a
generalization to $T^2$ of the well known $2^{nd}$ order Lax
differential operator $\partial_{z}^{2}+u_{2}(z)$ involved in the
analysis of the so-called KdV hierarchies and in $2d$ integrable
systems on the circle $S^1$. The second order example of eq.(48)
reads as
\begin{equation}
{}{\cal L}_{2}={\xi^{2}_{H}}+u_{2},
\end{equation}
and is suspected to describe, in a natural way, the analogue of
the KdV equation on $T^2$. Moreover, eq.(48) which is well defined
for $s\geq r\geq 0$ may be extended to negative integers by
introducing pseudo-differential operators of the type
$\xi^{-k}_{H}$, $k > 1$, whose action on the fields $u_{s}(z,
\omega)$ is given by the Leibnitz rule. It is now important to
precise how the operators ${\cal L}_{n}^{(r,s)}(u)$ act on
arbitrary functions $f(z, \omega)$ via the hamiltonian operators
$\xi^{k}_{H}$.\\\\ Striking resemblance with the standard case
\cite{GD1, GD2} leads us to write the following Leibnitz rules for
the local and non local differential operators in $\xi_{H}$
\begin{equation}
\xi_{H}^{n}{f}(z, \omega)=\sum_{s=0}^{n}c_{n}^{s}\{log H,
{f}\}^{(s)}\xi_{H}^{n-s},
\end{equation}
and
\begin{equation}
\xi_{H}^{-n}{f}(z, \omega)=\sum_{s=0}^{\infty
}(-)^{s}c_{n+s-1}^{s} \{log H, {f}\}^{(s)}\xi_{H}^{-n-s}
\end{equation}
where the $k^{th}$-order derivative $\{log H, {f}\}^{(k)}=
\underbrace{\{log H,\{log H,{...}\{log H, }\limits_{k
\hspace{0.1cm} times} f\}...\}\}$ on the torus $T^2$ is the
analogue of $f^{(k)}=\frac{\partial^{k} f}{\partial z^{k}}$, the
${k}$th derivative of $f$ in the standard case. Special examples
are given by
\begin{equation}
\begin{array}{lcl}
\xi_{H}\circ {f}&=&\{log H, f\}+f\xi_{H}\\\\
\xi_{H}^{2}\circ {f}&=&\{log H, f\}^{(2)}+2\{log H, f\}\xi_{H}+f\xi_{H}^{2}\\\\
\xi_{H}^{3}\circ {f}&=&\{log H, f\}^{(3)}+3\{log H,
f\}^{(2)}\xi_{H}+3\{log H,
f\}\xi_{H}^{2}+f\xi_{H}^{3}\\\\
\xi_{H}^{4}\circ {f}&=&\{log H, f\}^{(4)}+4\{log H,
f\}^{(3)}\xi_{H}+6\{log H, f\}^{(2)}\xi_{H}^{2}+4\{log H,
f\}\xi_{H}^{3}+f\xi_{H}^{4}\\\\
\end{array}
\end{equation}
and
\begin{equation}
\begin{array}{lcl}
\xi^{-1}_{H}\circ {f}&=&f\xi_{H}^{-1}-\{log H, f\}\xi^{-2}+\{log H, f\}^{(2)}\xi^{-3}-\{log H, f\}^{(3)}\xi^{-4}+..\\\\
\xi^{-2}_{H}\circ {f}&=&f\xi_{H}^{-2}-2\{log H, f\}\xi^{-3}+3\{log H, f\}^{(2)}\xi^{-4}+...\\\\
\xi^{-3}_{H}\circ {f}&=&f\xi_{H}^{-3}-3\{log H, f\}\xi^{-4}+...\\\\
\xi^{-4}_{H}\circ {f}&=&f\xi_{H}^{-4}+...\\\\
\end{array}
\end{equation}
As can be checked by using the Leibnitz rule, one have the
expected property
\begin{equation}
\xi_{H}^{n}.\xi_{H}^{-n}f(z, \omega)=f(z, \omega)
\end{equation}
A natural representation basis of non linear pseudo-differential
operators of spin $n$ reads as
\begin{equation}
{\mathcal P}_{n}^{(p,q)} \left[ u\right]
=\sum\limits_{i=p}^{q}u_{n-i} (z,\omega )\xi^{i}_{H}, \hspace{1cm}
p\leq q\leq -1
\end{equation}
This configuration, which is a direct extension of the local Lax
operators ${\cal L}_{n}^{(r,s)}(u)$ eq.(48), describes nonlocal
differential operators. Later on, we will use another
representation of pseudo-differential operators, namely, the
Volterra representation. The latter is convenient in the
derivation of the Gelfand-- Dickey (second hamiltonian structure)
of higher conformal spin integrable systems. Note by the way that
the non local Leibnitz rule eq.(51) is a special example of the
Volterra pseudo-operators as we will show in the forthcoming
sections.

\subsubsection{The algebra of differential operators
${\widehat\Sigma}$}

This is the algebra of differential operators of arbitrary spins
and arbitrary degrees. It's obtained from ${\widehat \Sigma}
_{n}^{(r,s)}$ by summing over all allowed degrees $(r, s)$ and
conformal weight (spin) $n$ in the following way
\begin{equation}
{\widehat{\Sigma }}=\oplus _{r\leq s}\oplus _{n\in Z}{\widehat{%
\Sigma }}_{n}^{(r,s)}.
\end{equation}
${\widehat{\Sigma }}$ is an infinite dimensional algebra which is
closed under the Lie bracket without any condition.\\
A remarkable property of this space is the possibility to split it
into six infinite dimensional classes of sub-algebras given by ${\widehat{%
\Sigma }}_{s}^{\pm },$ with $s=0,+,-$. These classes of algebras
are describing respectively the different values of the conformal
spin which can be zero, positive or negative. The $\pm $ upper
indices stand for the sign (positive or negative) of the degrees
quantum numbers. They are related to each other by conjugation of
the spin and degrees. Indeed, given two integers $s\geq r$, it's
not difficult to see that the spaces $\widehat{\Sigma }^{(r,s)}$
and ${\widehat{\Sigma }}^{(-1-r,-1-s)}$ are dual with respect to
the pairing product $(., .)$ defined as
\begin{equation}
({\mathcal L}^{(r,s)}, {\mathcal
P}^{(p,q)})=\delta_{1+r+q,0}\delta_{1+s+p,0}Res[{\mathcal
L}^{(r,s)}\circ{\mathcal P}^{(p,q)}]
\end{equation}
where the residue operation $Res$ is given by
\begin{equation}
Res[u_{j}\xi^{j}_{H}]=u_{-1}(z, \omega)
\end{equation}
As signaled previously, the residue operation $Res$ carries a
conformal weight $(-1, -1)$. \\
Note by the way that the $u_j$'s currents should satisfy a
conformal spin's duality with respect the following "scalar"
product
\begin{equation}
< u_{k}, u_{l}> =\delta_{k+l,1}\int{dz d\omega u_{1-k}(z,
\omega)u_{k}(z, \omega) }
\end{equation}
By virtue of this product the one dimensional spaces
${\widehat{\Sigma}}_{k}^{(0,0)}$ and
${\widehat{\Sigma}}_{1-k}^{(0,0)}$ are dual to each other. This
leads then to a splitting of the tensor algebra
${\widehat{\Sigma}}^{(0,0)}$ into two semi-infinite tensor
subalgebras ${\widehat{\Sigma}}_{+}^{(0,0)}$ and
${\widehat{\Sigma}}_{-}^{(0,0)}$, respectively characterized by
positive and negative conformal weights as shown here below
\begin{equation}
\begin{array}{lcl}
{\widehat{\Sigma}}_{+}^{(0,0)}&=&\oplus_{k>0}{\widehat{\Sigma}}_{k}^{(0,0)}\\\\
{\widehat{\Sigma}}_{-}^{(0,0)}&=&\oplus_{k>0}{\widehat{\Sigma}}_{1-k}^{(0,0)}
\end{array}
\end{equation}
We learn in particular that ${\widehat{\Sigma}}_{0}^{(0,0)}$ is
the dual of ${\widehat{\Sigma}}_{1}^{(0,0)}$. The previous
properties shows that the operation $< , >$ carries a conformal
weight $|<, >|= (-1, -1)$, consequently this is not a convenient
Lorentz scalar product. We need then to introduce a combined
Lorentz scalar product $<< , >>$ built out of $< , >$ eq.(59) and
the pairing product $( , )$ eq(57) such that $|<< ,
>>|=(0, 0)$ as follows
\begin{equation}
<< D_{m}^{(r, s)} , D_{n}^{(p,
q)}>>=\delta_{n+m,0}\delta_{1+r+q,0}\delta_{1+s+p, 0}\int {dz
Res[D_{m}^{(r, s)}\times D_{-m}^{(-1-s, -1-r)}]}
\end{equation}
With respect to this combined scalar product, one sees easily that
the subspaces ${\widehat{\Sigma}}_{++}, {\widehat{\Sigma}}_{0+}$
and ${\widehat{\Sigma}}_{-+}$ are dual to
${\widehat{\Sigma}}_{--}, {\widehat{\Sigma}}_{0-}$ and
${\widehat{\Sigma}}_{+-}$ respectively. The symbol
${\widehat{\Sigma}}_{+-}$ correspond for example to Lax operators
on $T^2$ with positive conformal spin and negative degrees while
${\widehat{\Sigma}}_{0+}$ corresponds to Lorentz scalar operators
of positive degrees. We conclude this section by making the
following remarks:\\
\textbf{1.} ${\widehat{\Sigma}}_{++}$ is the space of local
differential operators with positive definite conformal weight and
positive
degrees.\\
\textbf{2. }${\widehat{\Sigma}}_{--}$ is the Lie algebra of non
local differential operators with negative definite conformal
weight and negative degrees. \\
\textbf{3.} These two subalgebras ${\widehat{\Sigma}}_{\pm\pm}$
are used to built the Gelfand-- Dickey second hamiltonian
structure of integrable systems on the Torus $T^2$.

\subsection{$W_{n}$ symmetries from the ${{\Sigma}}_{++}\oplus
{{\Sigma}}_{--}$ algebras on $T^2$}
\subsubsection{The ${\Sigma}_{--}$ algebra in the Volterra basis}
The Lie algebra ${\Sigma}_{--}$ discussed previously is in fact
isomorphic to the maximal algebra of arbitrary negative definite
conformal spin and pure non local pseudo-differential operators
${\Sigma}_{-}^{(-\infty,-1)}$. In the Volterra representation a
generic elements of this algebra is given by
\begin{equation}
\Gamma^{(-\infty, -1)}(v)=\sum_{m=0}^{\infty}\alpha(m)
\Gamma_{-m}^{(-\infty, -1)}(v),
\end{equation}
where only a finite number of the coefficients $\alpha (m)$ is non
vanishing and where
\begin{equation}
\Gamma^{(-\infty,
-1)}_{-m}[v]=\sum_{j=1}^{\infty}\xi^{-j}_{H}\circ v_{j-m}(z,
\omega)
\end{equation}
More generally, given two integers $r$ and $s$ with $r\geq s\geq
0$, one can built new kinds of coset subalgebras ${
\Sigma}_{-}^{(-r, -s)}$ whose generators basis are given in the
Volterra representation as

\begin{equation}
\Gamma^{(-r, -s)}_{-m}[v]=\sum_{j=s}^{r}\xi^{-j}_{H}\circ
v_{j-m}(z, \omega)
\end{equation}
The number of independent fields $v_{j}(z, \omega)$ involved in
the above relation is equal to $(1+r-s)$ and corresponds then to
the dimension of ${\widehat \Sigma}_{-m}^{(-r,-s)}$. Note moreover
that with respect to the combined scalar product $<< ,
>>$ eq.(61), the subspace ${\Sigma}_{-}^{(-r,-s)}$ is nothing but the
dual of ${\Sigma}_{+}^{(s+1,r+1)}$.

Now, given two differential operators ${\mathcal L}^{(p,
q)}_{m}[u]$ and ${\Gamma}^{(-1-q, -1-p)}_{-n}[v]$ belonging to
${\widehat \Sigma}_{m}^{(p, q)}$ and
${\widehat\Sigma}_{-n}^{(-1-q, -1-p)}$ respectively, their residue
pairing reads as
\begin{equation}
Res[{\mathcal L}^{(p, q)}_{m}\circ {\Gamma}^{(-1-q,
-1-p)}_{-n}]=\sum_{i=p}^{q}u_{m-i}(z, \omega)v_{i+1-n}{(z,
\omega)}
\end{equation}
The conformal weight of the r.h.s. of the previous equation is
$(1+m-n, 1+m-n)$ where the values $m$ and $(-n)$ are the
contributions of the operators ${\mathcal L}^{(p, q)}_{m}[u]$ and
${\Gamma}^{(-1-q, -1-p)}_{-n}[v]$.

\subsubsection{The Gelfand-Dickey algebra on $T^2$}
We show in this section how one can derive the algebra of higher
spin currents from the so-called Gelfand-Dickey algebra on the
torus $T^2$. This can be done following the same lines of the
standard $sl_{n}$-GD algebra on the circle $S^1$ \cite{GD1, GD2}.
The first steps concerns the use of the algebra ${\Sigma}_{++}$
with the particular differential operators
\begin{equation}
{\cal L}_{n}^{(0, n)}[u]=\sum_{i=0}^{n}u_{n-i}\xi_{H}^{i}
\end{equation}
We consider then the $(n+1)$ fields $u_{j}(z, \omega)$ of
conformal weight $(j, j)$, $0\leq j\leq n$, involved in this
equation as the coordinates of a $(n+1)$ dimensional manifold
${\cal M}^{n+1}$. Let's denote by $\cal F$ the space of
differentiable functions on ${\cal M}^{n+1}$ of arbitrary positive
conformal spin. We have
\begin{equation}
{\cal F}=\oplus_{k\in N}{\cal F}_{k}
\end{equation}
Note that elements $ F_{k}$ of ${\cal F}_{k}$, for $k$ a positive
integer, depend on the $u_j(z, \omega)$'s and carry in general a
conformal spin index $k$
\begin{equation}
{F}_{k}[u] =F_{k}[u_{0}(z, \omega), ...,u_{n}(z, \omega)]
\end{equation}
The functional variation of these objects reads as
\begin{equation}
\delta {F}_{k}[u(z, \omega)] =\int
{dz}'{d\omega}'\sum_{j=0}^{n}\{\delta u_{j}(z',
{\omega}')\frac{\delta F_{k}[u(z, \omega)]}{\delta u_{j}(z',
{\omega}')}\}
\end{equation}
A special example consists in setting ${F}_{k}[u(z,
\omega)]=u_{k}(z, \omega)$, with $0\leq k\leq n$, we get
\begin{equation}
\delta {u}_{k}(z, \omega) =\int
{dz}'{d\omega}'\sum_{j=0}^{n}\{\delta_{jk}\delta(z-z')\delta(\omega-\omega')\delta
u_{j}(z', {\omega}')\}
\end{equation}
showing among other that
\begin{equation}
\frac{\delta F_{k}[u(z, \omega)]}{\delta u_{j}(z', {\omega}')}
\end{equation}
behaves as a $(k+1-j, k+1-j)$ conformal weight object. Since we
are focusing to derive $W_n$ symmetries with $(n-1)$ independent
conserved currents on the torus $T^2$ associated formally to an
$sl_n$-Lie algebra, one have to impose strong constraints on the
spin-$(0, 0)$ and spin-$(1, 1)$ fields $u_0$ and $u_1$ . This
leads then to a reduction of the $(n+1)$ dimensional manifold
${\cal M}^{(n+1)}$ to an $(n-1)$-dimensional submanifold ${\cal
M}^{(n-1)}$. Consistency gives
\begin{equation}
u_{0}={1}, \hspace{1cm} u_{1}=0
\end{equation}
The Lax operator eq.(66) reduces then to
\begin{equation}
{\cal L}_{n}^{(0,
n)}[u]=\xi_{H}^{n}+\sum_{i=0}^{n-2}u_{n-i}\xi_{H}^{i}
\end{equation}
Consider next the residue dual of the Lax differential operator
eq.(73). This is a pseudo-differential operator which has degrees
$(-1-n, 1)$ but unfixed conformal spin, say $(k-n, k)$ an integer.
These pseudo-operators reads in terms of the functions $F_{k}[u(z,
\omega)]$ as
\begin{equation}
\Gamma^{(-1-n, -1)}_{k-n}[F]=\sum_{j=1}^{n+1}\xi^{-j}_{H}\circ
v_{k+j-n}(z, \omega)
\end{equation}
where the $v_{k+j-n}(z, \omega)$'s with $1\leq j\leq n-1$ are
realized as
\begin{equation}
v_{k+j-n}(z, \omega)=\frac{\delta F_{k}[u(z, \omega)]}{\delta
u_{2+n-j}(z', {\omega}')}
\end{equation}
Note that the fields $v_{k+1}$ and $v_k$ define the residue
conjugates of $u_0$ and $u_1$ respectively. The explicit
determination of the pair of $v$'s fields requires the solving of
the following constraint equations
\begin{equation}
\begin{array}{lcl}
Res[{\cal L}^{(0, n)}_{n}\circ {\Gamma}^{(-1-n, -1)}_{k-n}] &=&
\sum\limits_{i=0}^{n-2}u_{n-i}(z, \omega)v_{i+k+1-n}{(z,
\omega)}\\ \\ \\
Res[{\cal L}^{(0, n)}_{n}\circ {\Gamma}^{(-1-n, -1)}_{1-n}[G]]&=&0
\end{array}
\end{equation}
Algebraic computations lead to
\begin{equation}
\begin{array}{lcl}
 v_{k+1}&=& 0 \\ \\
 v_k &=& \frac{1}{n}\sum\limits_{j=1}^{n}\sum\limits_{l=0}^{n-2}{\cal H}(j-i-1)(-1)^{j+1}c_{j}^{i}
 [u_{n-j}\times\frac{\delta {\cal F}_k}{\delta u_{n-i}}]^{(j-i-1)},
\end{array}
\end{equation}
as a solution of eqs.(76) respectively, where ${\cal H}(j)$ is the
Heaviside function defined as
\begin{equation}
{\cal H}(j)={\large \{}
\begin{tabular}{ll}
1, & if j $\geq$ 0 \\
0, & elsewhere
\end{tabular}
\end{equation}
Now given two functionals ${F}_{k}[u]$ and ${G}_{l}[u]$ depending
on the currents $u_{0}, ...,u_{n}$ with $u_{0}=1$ and $u_{1}=0$ as
well as corresponding pseudo-differential operators in the
Volterra representation ${\Gamma}_{k-n}^{(-1-n, -1)}[F]$ and
${\Gamma}_{l-n}^{(-1-n, -1)}[G]$. With these data, the
Gelfand-Dickey bracket reproducing the second hamiltonian
structure of integrable systems is given as follows
\begin{equation}
\{F_{k}[1], G_{l}[2]\}_{{\cal L}_{n}^{(0, n)}} =\int
{dz}{d\omega}Res[V_{k+n}^{(0, 2n-1)}({\cal L}, \Gamma)\circ
{\Gamma}_{1-n}^{(-1-n, -1)}]
\end{equation}
where we have used the following notation $F_{k}[u(z_1,
\omega_1)\equiv F_{k}[1]$ and $G_{l}[u(z_2, \omega_2)\equiv
G_{l}[2]$ with the definition
\begin{equation}
V_{k+n}^{(0, 2n-1)}({\cal L}, \Gamma)= {\cal L}_{n}^{(0,
n)}\circ(\Gamma_{k-n}^{(-1-n, -1)} \circ {\cal L}_{n}^{(0,
n)})_{+}-({\cal L}_{n}^{(0, n)}\circ \Gamma_{k-n}^{(-1-n,
-1)})_{+}\circ {\cal L}_{n}^{(0, n)}
\end{equation}
As usual, the suffix $(+)$ stands for the restriction to the local
part of the considered operation (positive powers of
$\xi_{H}^{j}$). One shows after a straightforward, but lengthy
computations, that the previous Gelfand-Dickey bracket reduces to
the following important form
\begin{equation}
\{F_{k}[1], G_{l}[2]\}_{GD}=\sum_{i, j= 0}^{n-2}\int {dz}{d\omega}
[ \frac{\delta {\cal F}_{k}[1]}{\delta u_{n-i}(z', \omega')} {\cal
D}(n, i, j, z', \omega', u) \frac{\delta {\cal G}_{l}[2]}{\delta
u_{n-j}(z', \omega')} ]
\end{equation}
where the operator ${\cal D}(n, i, j, z', \omega', u)$ is a
nonlinear local differential operator of conformal weight $|{\cal
D}|=(2n-1-i-j, 2n-1-i-j)$. The knowledge of this operator is a
central steps towards computing the Gelfand-Dickey bracket. It's
explicit determination is a tedious work and we will avoid the
presentation of all our calculus and restrict our self to the
global forms leading to the extended higher conformal spins
symmetries on the torus $T^2$.\\\\
First of all, we have to underline the important situation for
which we can set $F_k=u_k$ and $G_l=u_l$ with $2\leq k, l\leq n$.
The basic Gelfand-Dickey brackets associated to the higher spin
symmetries (conformal and W-extensions) expressed in terms of the
canonical fields $u_j, 2\leq j\leq n$ are given as follows
\begin{equation}
\{u_{k}[1], u_{l}[2]\}_{GD}={\cal D}(n, n-k, n-l,
u)\delta(z_1-z_2)\delta(\omega_{1}-\omega_{2})
\end{equation}
It's easily seen that ${\cal D}(n, n-k, n-l, u)$ is a nonlinear
differential operator of conformal weight $|{\cal D}(n, n-k, n-l,
u)|=(k+l-1, k+l-1)$. The particular examples that we are
interested in concern the conformal and $W_3$ symmetries
associated respectively to the restriction of the order of the Lax
operators to $n=2$ and $3$. For $n=2$, we have $k=l=2$, the unique
Poisson bracket corresponding the the classical version of the
conformal symmetry in $T^2$ is given by
\begin{equation}
\{u_{2}[1], u_{2}[2]\}_{GD}={\cal D}(2, 0, 0,
u)\delta(z_1-z_2)\delta(\omega_{1}-\omega_{2})
\end{equation}
and the differential operator ${\cal D}(2, 0, 0, u)$ takes the
following form
\begin{equation}
{\cal D}(2, 0, 0, u)=\frac{1}{2}\xi_{H}^{3}+2u_{2}(z,
\omega)\xi_{H}+\{log H, u_{2}\}
\end{equation}
This is a third order differential operator extending the one
appearing in the classical version of the Virasoro symmetry. For
the nearest values of $z$ and $z'$, the conformal algebra
eqs.(83-84) reads as
\begin{equation}
\{u_{2}[1], u_{2}[2]\}_{GD}=[u_{2}(z, \omega)+u_{2}(z',
\omega')]\{log H, \delta(z-z')\delta(\omega-\omega')\}+
\frac{1}{2}\{log H, \delta(z-z')\delta(\omega-\omega')\}^{(3)}
\end{equation}
where $\{log H, \delta(z-z')\delta(\omega-\omega')\}^{(3)}=\{log
H, \{log H, \{log H, \delta(z-z')\delta(\omega-\omega')\}\}\}$ is
the third order derivation of
$\delta(z-z')\delta(\omega-\omega')$. eq.(85) corresponds then to
the classical version of the Virasoro algebra on $T^2$. Remark
that the central charge corresponding to this algebra is $c=6$,
this is inherited from the simple choice of $u_0=1$. If one
choices the Lax operator to be
\begin{equation}
{\cal L}_{n}^{(0,
n)}[u]=\frac{c}{6}\xi_{H}^{n}+\sum_{i=0}^{n-2}u_{n-i}\xi_{H}^{i},
\hspace{1cm} {with}\hspace{1cm} u_0= \frac{c}{6}
\end{equation}
in this case the bianalytic Virasoro algebra becomes
\begin{equation}
\{u_{2}[1], u_{2}[2]\}=[u_{2}+u_{2'}]\{log H,
\delta(z-z')\delta(\omega-\omega')\}+ \frac{c}{12}\{log H,
\delta(z-z')\delta(\omega-\omega')\}^{(3)}
\end{equation}

Similarly, the $W_{3}$-extension of the Frappat et al. symmetry on
$T^{2}$ is obtained from the generalized Gelfand-Dickey bracket
eq(82) and gives for $n=3$ three brackets associated to the fields
$u_2$ and $u_3$ as follows
\begin{equation}
\begin{array}{lcl}
\{u_{2}[1], u_{2}[2]\}_{GD}&=&{\cal D}(3, 1, 1,
u)\delta(z_1-z_2)\delta(\omega_{1}-\omega_{2})\\\\
\{u_{2}[1], u_{3}[2]\}_{GD}&=&{\cal D}(3, 1, 0,
u)\delta(z_1-z_2)\delta(\omega_{1}-\omega_{2})\\\\
\{u_{3}[1], u_{3}[2]\}_{GD}&=&{\cal D}(3, 0, 0,
u)\delta(z_1-z_2)\delta(\omega_{1}-\omega_{2})
\end{array}
\end{equation}
where
\begin{equation}
\begin{array}{lcl}
{\cal D}(3, 1, 1,u)&=& 2\xi_{H}^{3}+2u_{2}\xi_{H}+\{log H, u_{2}\}\\\\
{\cal D}(3, 1, 0,u)&=& -\xi_{H}^{4}-u_{2}\xi_{H}^{2}+(3u_3-2\{log H, u_{2}\})\xi_{H}+2\{log H, u_3\}-\{log H, u_2\}^{(2)},\\\\
{\cal D}(3, 0, 0,u)&=&
-\frac{2}{3}\xi_{H}^{5}-\frac{4}{3}u_{2}\xi_{H}^{3}-3\{log H,
u_{2}\}\xi_{H}^{2}+ (2\{log H, u_{3}\}-2\{log H,
u_{2}\}^{(2)}-\frac{2}{3}u_{2}u_{2})\xi_{H}\\\\&+&(\{log H,
u_3\}^{(2)}-\frac{2}{3}\{log H, u_2\}^{(3)}-\frac{2}{3}u_{2}\{log
H, u_2\})

\end{array}
\end{equation}

\newpage
\section{Concluding Remarks}
We presented in this paper some important aspects of higher
conformal spin symmetries on the bidimensional torus $T^2$. These
symmetries, generalizing the Frappat et al. conformal symmetries
by adding currents of conformal spin 3 in a non standard way, are
also shown to be derived, in their semi-classical form, from the
Gelfand-Dickey bracket that we are computing explicitly. \\\\
We underline at this level that the obtained $W_3$ extension of
the Frappat et al. conformal symmetry on $T^2$ exhibits many
remarkable features. The first one concerns the introduction of
new kind of derivation inherited from diff$(T^2)$ and that takes
the following form $\xi_{H}\equiv \{logH, .\}=\partial_{z} log H
\partial_{\omega}-\partial_{\omega} log H \partial_{z}$ for arbitrary bianalytic function $H(z,
\omega)$. These logarithmic derivatives, corresponding to
Hamiltonian vector fields, are central in the present study as
they are assuring a consistent description of diff$(T^2)$
symmetries. Besides it's diff$(T^2)$ invariance origin, the vector
field $\xi_{H}$ is very useful as it can join in a compact form
the bi-complex derivatives $\partial_{z}$ and $\partial_{\omega}$.
Note that for functions $H$ living on a $n$-dimensional torus
$T^n$, the logarithmic derivative can be written as
\begin{equation}
\{log H , \} =\sum\limits_{i,j=1}^{n}\Omega ^{ij}
\partial _{i} \log {H} \partial _{j}
\end{equation}
where $H=H(z_{1} ,...,z_{n} ,\omega _{1} ,...,\omega _{n})$ and
$\Omega$$_{ij}$ = - $\Omega$$_{ji}$ is the usual $n \times n $
antisymmetric matrix.\\\\
The second crucial remark concerning the derived Gelfand-Dickey
algebra is that the $nth$ order local differential operator used
is of the form
\begin{equation}
{\cal L}_{n}^{(0,
n)}[u]=\xi_{H}^{n}+\sum_{i=0}^{n-2}u_{n-i}\xi_{H}^{i}
\end{equation}
This is a Lax operator belongings formally to the $A$-series of
simple Lie algebra and having $(n-1)$ coordinates functions
$\{u_{k}$, $ k=2,3,...n\}$, where we have set $u_{0}=1$ and
$u_{1}=0$. This is a natural generalization of the well known
differential $sl_{2}$-Lax operator ${\cal
L}_{KdV}={\xi^{2}_{H}}+u_{2}$ associated to the diff$(T^2)$-KdV
integrable hierarchy that we will discuss later \cite{sed}.

We have to underline that the $sl_{n}$-Lax operators play a
central role in the study of integrable models and more
particularly in deriving higher conformal spin algebras
($W_{k}$-algebras) from the extended Gelfand-Dickey second
Hamiltonian structure. Since they are also important in recovering
$2d$ conformal field theories via the Miura transformation, we are
convinced about the possibility to extend this property, in a
natural way, to diff$(T^2)$ and consider the $T^2$-analogues of
the well known $2d$ conformal models namely: the
$sl_{2}$-Liouville field theory and its $sl_{n}$-Toda extensions
and also the Wess-Zumino-Novikov-Witten conformal model. For
instance, consider the KdV Lax operator that we can write by
virtue of the Miura transformation as
\begin{equation}
{\cal L}_{KdV}= \xi_{H}^{2}+u_{2}(z, \omega)=(\xi_{H}+\{log H,
\phi \})\times (\xi_{H}-\{log H, \phi \})
\end{equation}
where $\phi $ is a Lorentz scalar field. As a result we have
\begin{equation}
u_{2}=-\{log H, \phi \}^{(2)}-\{log H, \phi \}^2
\end{equation}
which should describe the classical version of the stress energy
momentum tensor of conformal field theory on the torus $T^2$.
Using $\mathbf{C}\times \mathbf{C}$ bicomplex coordinates
language, we can write
\begin{equation}
{\mathcal{T}}(z, \omega)\equiv u_{2}(z, \omega)=-\{log H, \phi
\}^{(2)}-\{log H, \phi \}^2
\end{equation}
where $\{log H, \phi \}^{(2)}$ is the second order derivative of
the Lorentz scalar field $\phi$ with respect to the $T^2$
symmetry. The conservation for this bianalytic conformal current
${\mathcal T}(z, \omega)$, leads to write the following
differential equation
\begin{equation}
\{log {\bar{K}}, \{log H, \phi \}\}=e^{2\phi}
\end{equation}
where ${\bar K}=K(\bar z, \bar \omega)$ is an arbitrary bianalytic
function of $\bar z$ and $\bar \omega$ carrying in general an
$({\bar n}_{0}, {\bar m}_{0})$ $U(1)\times U(1)$ charge. Note also
that ${\bar K}$ is not necessarily the complex conjugate of the
function $H$ considered earlier. Our experience with conformal
field theory and integrable systems leads to conclude that the
later "second order" differential equation is nothing but the
conformal Liouville like equation of motion on the Torus $T^2$.
This equation of motion is known to appear in this context as a
compatibility relation with the conservation of the stress energy
momentum tensor ${\mathcal T}(z, \omega)$ namely $\{log {\bar{K}},
{\mathcal T}(z, \omega)\}=0$ or equivalently $\{log {\bar{K}},
\{log H, \phi \}^{(2)}\}+2\{log H, \phi\}\{log {\bar{K}}, \{log H,
\phi \}\}=0$.\\\\
We end this discussion by noting the possibility to interpret the
obtained diff$(T^2)$-Gelfand-Dickey symmetries eqs() corresponding
to the $nth$-order Lax differential operator eq.(91) as been the
classical form of the $W_n$ symmetries behind the $\{sl_n\}$-Toda
like conformal field theory, generalizing the diff$(S^1)$ Toda
field theory \cite{Toda} and whose action is assumed to have the
following form
\begin{equation}
S[\phi]=\int d^{2}{z}d^{2}{\omega}\{\frac{1}{2}\{log H,
\phi\}.\{log {\bar K},
\phi\}-(\frac{m}{\beta})^{2}\sum\limits_{i=1}^{n-1}exp(\beta\alpha_{i}.\phi)\}
\end{equation}
In this bosonic Toda field theory on $T^2$, the Cartan
subalgebra-valued scalar field $\phi$ is given by
$\phi=\sum\limits_{i=1}^{n-1}\alpha_{i}\phi_i$ where $\phi_i$ is
an $n-$component scalar field and $\{\alpha_{i}$,
$i=1,2,...,n-1\}$ are $(n-1)$-simple roots for the underlying Lie
algebra ${\mathcal{G}}$. the parameters $m$ and $\beta$ are
coupling constants. The derivatives of the scalar field $\phi$ in
the kinetic term are $\{log H, \phi\}$ and $\{log {\tilde H},
\phi\}$ giving then the extension of diff$(S^1)$-derivatives
$\partial_{z} \phi$ and $\bar{\partial_{z}} \phi$. The equations
of motion that we can derive from the Toda action eq.(96) are
summarized as follows
\begin{equation}
\beta\{log {\bar K}, \{log H,
\phi\}\}+m^2\sum\limits_{i=1}^{n-1}\alpha_{i}exp(\beta\alpha_{i}.\phi)\}=0
\end{equation}
A particular example is given by the $\{sl_2\}$ conformal
Liouville model eq.(95).
\\\\
{\bf{Acknowledgments}}\newline {\small I would like to thank the
Abdus Salam International Center for Theoretical Physics (ICTP)
for hospitality. I present special thanks to the high energy
section and to its head Prof. Seif Randjbar-Daemi for considerable
scientific help. Best acknowledgements are presented to the office
of associates for the invitation and scientific helps. I
acknowledge the valuable contributions of OEA-ICTP in the context
of NET-62 program}


\begin{thebibliography}{99}
\bibitem{CFT}
A.~A.~Belavin, A.~M.~Polyakov, A.~B.~Zamolodchikov,
Nucl.\ Phys.\ B {\bf 241}, 333 (1984).

V.~S.~Dotsenko and V.~A.~Fateev,
Nucl.\ Phys.\ B {\bf 240}, 312 (1984).

P.~H.~Ginsparg,
Les houches Lectures (1988)

\bibitem{string} M. Green, J. Schwarz, and E. Witten, Superstring
Theory, Cambridge, 1986. \

S.~Randjbar- Daemi and J.~A.~Strathdee, ``Introductory lectures on
CFT and Strings,''
{\it  In Trieste 1989, Proceedings, High energy physics and
cosmology 2-79.}

\bibitem{stat} C. Itzykson, H. Saluer, J.B. Zuber (eds.): conformal invaraince
and applications to statistical machanics. Singapore: World
scientific 1988\\
J. Mussardo, Phys. Rep. C 218, 275-382 (1992.

\bibitem{int} L.D.Faddeev, L.A.Takhtajan, Hamiltonian methods and the
theory of solitons, 1987, \\
 E. Date, M. Kashiwara, M. Jimbo and T. Miwa
In "Nonlinear Integrable Systems", eds. M. Jimbo and T. Miwa, World Scientific (1983),\\
 A. Das, Integrable Models, World scientific, 1989.

\bibitem{zamo1} A. B. Zamolodchikov, Theor. Math. Phys. 65 (1985) 1205;\\ V. A.
Fateev and A. B. Zamolodchikov, Nucl. Phys. B304 (1988) 348;

\bibitem{zamo2}
  P.~Di Francesco, C.~Itzykson and J.~B.~Zuber,
  Commun.\ Math.\ Phys.\ {\bf 140}, 543 (1991).

P. Bouwknegt and K.Schoutens, Phys. Rep. 223 (1993) 183;

\bibitem{Wgrav} K. Schoutens, A. Servin, and P. Van Nieuvenhuizen, Phys. Len. B 243, 245 (1990);
E. Bergshoeff, A. Bilal, and K. S. Stelle, TH 5924/90;

\bibitem{f1} V. A. Fateev and S. Lukyanov, Int. J. Mod. Phys. A 3, SO7
(1987);\\E Bais, P. Bouwknegt, M. Surridge, and K. Schoutens,
Nucl. Phys.B 304, 348 (1988);\\ L. Romans, Nucl. Phys.B 352, 829
(1991);\\ E. Bergshoeff, C.N.Pope, L.J.Romans, E.Sezgin, X.Shen,
Phys.Lett.B 245,447 (1990).

\bibitem{f2} V. G. Drintiel'd and V. V. Sokolov, Sov. J. Math. 38, 1975 (1985).

\bibitem{f3}
  E.~H.~Saidi and M.~B.~Sedra,
  Class.\ Quant.\ Grav.\ {\bf 10}, 1937 (1993);

  Int.\ J.\ Mod.\ Phys.\ A {\bf 9}, 891 (1994).

\bibitem{lie}J. Humphreys, 
(Springer-Verlag, Berlin, 1972).

  V.~G.~Drinfeld and V.~V.~Sokolov,
  J.\ Sov.\ Math.\ {\bf 30}, 1975 (1984).

T. Inami and H. Kamro, Commun. Math. Phys 136, 519 (1991);\\ Nucl.
Phys. B 359, 201,(1991);

\bibitem{ga} A. Bilal, V. V. Fock, and I.
I. Kogan, Nucl. Phys. B 359, 635 (1991);\\ M. Bershadsky and H.
Ooguri, Commun. Math. Phys. 126,49 (1989);\\ G. M. Stotkov, M.
Stanishkov, and C. J. Zhu, Nucl. Phys. B 356, 439 (1991).\\E. H.
Saidi and M. Zakkari, Phys. Lett. B 281, 507 (1992).

\bibitem{GD}
  J.~L.~Gervais,
  Phys.\ Lett.\ B {\bf 160}, 277 (1985).

  A.~Bilal and J.~L.~Gervais,
  Phys.\ Lett.\ B {\bf 206}, 412 (1988).

  I.~Bakas,
  Nucl.\ Phys.\ B {\bf 302}, 189 (1988).

  P.~Mathieu,
  Phys.\ Lett.\ B {\bf 208}, 101 (1988).

  K.~Yamagishi,
  Phys.\ Lett.\ B {\bf 259}, 436 (1991).

\bibitem{GD1}
  I.~Bakas,
  Commun.\ Math.\ Phys.\  {\bf 123}, 627 (1989).

\bibitem{GD2}
  E.~H.~Saidi and M.~B.~Sedra,
  J.\ Math.\ Phys.\ {\bf 35}, 3190 (1994).

\bibitem{frap} L. Frappat E.Ragoucy, P.Sorba, F.Thuiller, H.H$\emptyset$gasen, Nucl. Phys. B334 (1990) 250.

\bibitem{anton} I. Antoniadis, P. Ditsas, E. Floratos and J. Iliopoulos, Nucl.
Phys. B300 (1988) 549.

\bibitem{sss}
  E.~H.~Saidi, M.~B.~Sedra and A.~Serhani,
  Phys.\ Lett.\ B {\bf 353}, 209 (1995);

  Mod.\ Phys.\ Lett.\ A {\bf 10}, 2455 (1995).

\bibitem{frac}
  E.~H.~Saidi, J.~Zerouaoui and M.~B.~Sedra,
  Class.\ Quant.\ Grav.\ {\bf 12}, 1567 (1995);

  Class.\ Quant.\ Grav.\ {\bf 12}, 2705 (1995).

\bibitem{arak} T. A. Arakelyan and G. K. Savvidy, Phys. Lett. B214 (1988)
350.

\bibitem{Toda} K. Toda and S.J. Yu, J. Math. Phys. 41 (2000)
4747; J. Nonlinear Math. Phys. Suppl. 8(2001)272; Inverse Problems
17 (2001) 1053.

\bibitem{sed}
M.B. Sedra, Integrable KdV hierarchies on the Torus $T^2$, work in
progress.

\bibitem{Toda}
  P.~Mansfield,
  Nucl.\ Phys.\  B {\bf 208}, 277 (1982).

  P.~Mansfield,
  Nucl.\ Phys.\  B {\bf 222}, 419 (1983).

  J.~Evans and T.~J.~Hollowood,
  Nucl.\ Phys.\  B {\bf 352}, 723 (1991),[B {\bf 382}, 662 (1992)].


\end{thebibliography}
\end{document}